\documentclass[11pt,a4paper]{article}
\pdfoutput=1
\usepackage{jheppub}
\usepackage{amsmath}
\usepackage{verbatim}
\usepackage{graphicx}
\usepackage{mathrsfs}
\usepackage{appendix}
\usepackage{caption}
\usepackage{float}
\usepackage{subfig}
\usepackage{mathtools}
\usepackage{mathtools}
\usepackage{caption}
\usepackage{amssymb}
\usepackage{color}
\usepackage{inputenc, array}
\usepackage{textcomp}
\usepackage[dvipsnames]{xcolor}

\newcommand{\e}{\epsilon}
\newcommand{\bL}{\bar{L}}

\newcommand{\z}{{\bar z}}

\newcommand{\be}[1]{ \begin{equation}\label{#1} }
\newcommand{\ee}{\end{equation}}
\newcommand{\bea}[1]{\begin{eqnarray}\label{#1} }
\newcommand{\eea}{\end{eqnarray}}
\newcommand{\bes}{\begin{subequations}}
\newcommand{\ees}{\end{subequations}}

\newcommand{\p}{\partial}

\newcommand{\refb}[1]{(\ref{#1})}

\DeclarePairedDelimiterX\braket[2]{\langle}{\rangle}{#1 \delimsize\vert #2}

\title{3D Stress Tensor for Gravity in 4D Flat Spacetime}

\author[a]{Arjun Bagchi,} \author[a]{Prateksh Dhivakar,} \author[b]{and Sudipta Dutta.} \author{\\}

\affiliation[a]{Indian Institute of Technology Kanpur, Kalyanpur, Kanpur 208016. INDIA. \\}

\affiliation[b]{Physique théorique et mathématique, Université Libre de Bruxelles,
	Campus Plaine - CP 231, 1050 Bruxelles, Belgium. \\}

\emailAdd{abagchi@iitk.ac.in, prateksh@iitk.ac.in, sudipta.dutta@ulb.be}

\abstract{Three dimensional (3d) Carrollian CFTs are potential co-dimension one holographic duals of 4d asymptotically flat spacetimes that live on the whole of the null boundary. In this paper, we show that the {\em local stress tensor} of the 3d Carrollian conformal theory (without any additional sources) constructed in terms of the geometric structure at asymptotic null infinity naturally encodes {\em both} the leading and subleading soft graviton theorems. We relate the 3d Carroll stress tensor to the 2d Celestial one and show how the 3d version naturally localises the non-local 2d Celestial stress tensor. We also comment on the relation with  stress tensors in relativistic 3d CFTs and connections to the flat limit of AdS/CFT.}

\preprint{}

\begin{document}
\maketitle

\section{Introduction}

The very rich structure of infrared physics in asymptotically flat spacetimes (AFS) has become the cynosure of recent interest and a rather remarkable story has emerged relating asymptotic symmetries, gravitational memory effects and soft theorems in an intricate web of relations \cite{Strominger:2013jfa,Strominger:2017zoo,Pasterski:2021rjz,Raclariu:2021zjz}. 

\subsection*{Asymptotic symmetries and soft theorems}

The asymptotic symmetries algebra of four dimensional (4d) AFS at its null boundary ($\mathscr{I}^\pm$) is given by the Bondi-van der Burg-Metzner-Sachs (BMS$_4$) algebra \cite{bondibms,PhysRev.128.2851}:
\begin{subequations}\label{bms4}
\begin{align}
[L_n, L_m] &= (n-m)L_{m+n}, \quad [L_n, M_{r,s}] = \left(\frac{n-1}{2} -r\right) M_{n+r, s}, \\
[\bL_n, \bL_m] &= (n-m)\bL_{m+n},  \quad [\bL_n, M_{r,s}] = \left(\frac{n-1}{2} -s\right) M_{r, n+s}, \\
[M_{r,s}, M_{p,q}] &=0.
\end{align}
\end{subequations}
The structure of null infinity in 4d is $R\times S^2$ where $R$ is a null line. In the above, $M_{r,s}$ are the generators of angle-dependent translations of the null direction called supertranslations. $L_n$ and $\bL_n$ generate super-rotations, which are enhanced from the usual Lorentz algebra to two copies of the Witt algebra \cite{Barnich:2009se} \footnote{There are more general enhancements to the BMS algebra possible which includes the superrotations enhancing to the full diffeomorphisms of the sphere \cite{Campiglia:2014yka}. We will not be interested in these extensions in this work.}.  

Soft theorems are statements about the infrared (IR) of scattering amplitudes involving massless particles. 
The low energy expansion of scattering amplitudes involving a massless particle can be written in terms of the scattering amplitude without the soft particle and a multiplicative factor. This multiplicative factor has universal behaviour at the leading and (sometimes) sub-leading order. This was first studied by Low \cite{Low:1954kd,PhysRev.110.974} in the context of quantum electrodynamics (QED) and later on generalised to gravity by Weinberg \cite{PhysRev.140.B516}. These phenomena have come to be known as the leading soft theorems. Although the sub-leading soft theorem was known in QED previously \cite{PhysRevLett.20.86}, its existence in gravity was discovered only recently in \cite{Cachazo:2014fwa}. 

We will be interested in the soft graviton theorems in this work. Let us consider a scattering amplitude $\mathcal{M}_{n+1}$ of $(n+1)$ particles involving one graviton. Using Feynman diagrammatics one can show that this amplitude admits the following expansion as the energy of the graviton tends to zero \cite{PhysRev.140.B516,Cachazo:2014fwa}:
\begin{subequations}\label{gravisoft}
\begin{align} 
	\mathcal{M}_{n+1}(p_i,k)&=\left[\frac{1}{\omega}\sum_{i=1}^{n}\frac{\epsilon_{\mu\nu}p_i^\mu p_i^\nu}{p_i.\hat{k}}+\sum_{i=1}^{n}\frac{\epsilon_{\mu\nu}p_i^\mu k_{\lambda}}{p_i.\hat{k}}\hat{J}^{\nu\lambda}_i+\mathcal{O}(\omega)\right]\mathcal{M}_n(p_i)\\ &= \left[\frac{1}{\omega}S_0+ S_1 + \mathcal{O}(\omega)\right] \mathcal{M}_n(p_i). 
\end{align}
\end{subequations}
In the above expression $p_i$ and $k$ respectively denote the momentum of hard and soft (the zero energy graviton) particles. $\omega$  and $\epsilon_{\mu \nu}$ are the energy and polarisation vector of  the graviton. $\hat{J}^{\nu\lambda}_i$ is the angular momentum operator acting on the $i$th particle. The coefficients $S_0$ and $S_1$ are universal in nature and known respectively as the Weinberg and the Cachazo-Strominger soft factors. 

The connection between asymptotic BMS symmetries and the soft theorems manifests itself rather beautifully. It turns out that the soft theorems can be understood as BMS Ward identities \cite{Strominger:2014pwa,He:2014laa}. The leading Weinberg soft theorem is a consequence of supertranslation invariance \cite{He:2014laa} and the super-rotation Ward identity leads to the sub-leading soft graviton theorem \cite{Kapec:2016jld}. 

\subsection*{Flat Holography: Celestial and Carrollian}
Hand in hand with the developments in the IR physics in flat spacetimes, there are renewed attempts to understand holography in AFS. Recent development has proceeded in two principle directions called Celestial and Carrollian holography. The more popular approach, called Celestial holography, is a co-dimension two approach where the holographic dual to 4d AFS is proposed to be a 2d relativistic conformal field theory that lives on the celestial sphere at null infinity \cite{Strominger:2017zoo}. This approach has led to a lot of new insights into scattering amplitudes and asymptotic symmetries \cite{Strominger:2017zoo,Pasterski:2021rjz,Raclariu:2021zjz}. The Celestial approach makes the Lorentz symmetry manifest while somewhat obscuring the action of bulk translations. The principle observation is that the 4d Lorentz group in the bulk acts like the global 2d conformal group on the boundary. When written in terms of basis that diagonalises boosts \cite{Pasterski:2016qvg,Pasterski:2017kqt}, the link between the boundary and bulk becomes clearer. An integral transformation, viz. a Mellin transformation \cite{Pasterski:2016qvg}, on the 4d S-matrix elements maps them to correlation functions in the dual 2d CFT. This has connection has been exploited to find out properties of the dual field theory from known physics in the bulk. 

An earlier proposal of a holographic dual in terms of a non-Lorentzian CFT \cite{Bagchi:2010eg,Bagchi:2012cy,Duval:2014uva,Bagchi:2016bcd}, now better known as Carrollian holography, has had some success in terms of a correspondence between gravity in 3d flat spacetimes and a 2d Carrollian CFT \cite{Bagchi:2012xr,Barnich:2012xq,Bagchi:2014iea,Bagchi:2015wna,Jiang:2017ecm}. A body of recent work has emerged where the connection between 3d Carrollian CFTs and bulk physics in 4d asymptotically flat spacetimes has been explored \cite{Bagchi:2022emh,Donnay:2022aba,Donnay:2022wvx,Dutta:2022vkg,Bagchi:2023fbj,Salzer:2023jqv,Saha:2023hsl,Bagchi:2023cen,Mason:2023mti,Nguyen:2023vfz,Have:2024dff,Alday:2024yyj,Kraus:2024gso,Bagchi:2024efs,Bekaert:2024itn,Liu:2024llk} and in particular, in parallel to the connection between 3d Carrollian correlation functions and 4d S-matrix elements has been established via the so-called {\em Modified Mellin} transformation \cite{Banerjee:2018gce,Bagchi:2022emh}, which as the name suggests is a modification of the original Mellin transform used for the Celestial case. The modified Mellin transformation has the advantage that it renders graviton amplitudes in 4d bulk finite, unlike the usual Mellin transform \cite{Banerjee:2019prz}. 

Carrollian holography is co-dimension one, much like the original AdS/CFT correspondence \cite{Maldacena:1997re} and has the advantage that bulk translations now act naturally, without shifting weights of boundary operators like in the Celestial case \cite{Bagchi:2023cen}. If (some aspects of) flatspace holography in D dimensions were to emerge from a limit from AdS$_{D+1}$/CFT$_D$, it is also natural that the boundary theory does not reduce in dimension in the limit, but the symmetry algebra governing the theory contracts from $so(D,2)$ to $iso(D,1)$ to match the isometries of the bulk \cite{Bagchi:2016bcd,Bagchi:2019xfx}.   

\subsection*{Stress tensors}
In AdS/CFT, the stress tensor of the boundary theory is dual to metric fluctuations in the bulk \cite{Witten:1998qj,Balasubramanian:1999re}. Correlation functions of stress tensors are related to scattering of gravitons. Of course in AdS, due to the timelike nature of the asymptotic boundary, the notion of scattering is changed since there are no asymptotic ``in'' and ``out'' states. But one can create (or annihilate) particles by changing boundary conditions at infinity and scattering amplitudes are defined by transition amplitudes between these states \cite{Balasubramanian:1999ri,Giddings:1999qu,Penedones:2010ue,Fitzpatrick:2011jn}. In the flat space limit, the AdS scattering amplitudes go over to flat space scattering amplitudes \cite{Penedones:2010ue,Fitzpatrick:2011jn}. 
The large radius limit that sends AdS to flatspace is a Carroll limit on the boundary. Thus it is very tempting to presume that a properly defined Carrollian stress tensor would capture the details of scattering of gravitons in flat spacetime. 

As we reviewed above, soft graviton theorems are related to BMS Ward identities. From the point of view of the 2d Celestial CFT, the stress tensor of the theory captures the sub-leading soft graviton theorem while one has to postulate the existence of a spin-one current in the theory to account for the leading Weinberg soft theorem. This is somewhat unsatisfactory, since if (some aspects of) flat holography were to arise from a consistent limit of AdS/CFT, presumably both the leading and subleading soft theorems should be related to stress tensor Ward identities in the dual theory. 

In this paper, we show that by considering a 3d Carrollian CFT, instead of a 2d Celestial CFT, we are able to fulfil this expectation. Both soft graviton theorems indeed are packaged into the 3d Carroll stress tensor Ward identities. The 2d Celestial stress tensor is also known to have issues with locality. We comment on how the 3d Carroll stress tensor helps alleviate such concerns. 

In contrast with \cite{Donnay:2022aba,Donnay:2022wvx}, our construction does not necessitate the inclusion of sources. The theory of gravity in 4d AFS is a unitary theory. So a  holographic dual field theory to 4d AFS also should be unitary. Although a Carroll CFT living purely on $\mathscr{I}^+$ would be non-unitary since there is radiation leaking out of the future null boundary, all this radiation comes from $\mathscr{I}^-$. A Carroll CFT which takes both boundaries into account would be unitary. We believe our construction is implicitly a Carroll CFT which encodes both $\mathscr{I}^\pm$ and is unitary. 
Our paper would draw heavily on the methods of the modified Mellin transformation \cite{Banerjee:2018gce} and the construction of the stress tensor in \cite{Dutta:2022vkg}. Other relevant work on boundary Carroll stress tensors include \cite{Mann:2005yr,Bagchi:2015wna,Ciambelli:2018ojf,Ciambelli:2018wre,Chandrasekaran:2021hxc,deBoer:2021jej,Freidel:2022vjq,Saha:2023hsl,Adami:2024rkr}

\subsection*{Outline of the paper}

As previously addressed, the main purpose of this work is to investigate the implications of soft theorems in a putative codimension one Carrollian dual. In AdS/CFT correspondence the bulk metric certifies the existence of a conformal stress tensor in the boundary theory. Now as the boundary of asymptoticlly flat spacetime is a null hypersurface, in similar lines, one sould expect the existence of a Carrollian stress tensor. In this paper we initiate an investigation in this direction and partially establish such result using hints from the soft graviton theorems.

We start by introducing a new basis that rewrite the flat space scattering amplitudes as Carrollian correlation functions in \S 2. We express the leading and subleading soft graviton theorems in this basis. Then in \S 3, we shift our focus in the construction of 3d conformal Carroll stress tensors. We define the stress tensors geometrically using the variation of intrinsic Carroll structures on null infinity. Using these stress tensors we then derive the BMS currents and charges in dual field theory. We provide the transformation rules of Carrollian primary fields with respect to these infinite symmetries and finally derive the Ward identities of conformal Carroll stress tensors. In \S 4 we state our main result where we identify these stress tensor Ward identities with the soft graviton theorems in Carrollian basis. This identification leads us to a relation between soft gravitons and boundary stress tensor components. In \S 5 we compare and relate our results with the framework of celestial holography. In the celestial proposal these soft graviton theorems indicate the existence of a Virasoro (2d-CFT) stress tensor and a U(1) Kac-Moody current. We discuss how these operators are related with the Carrollian stress tensors. Finally \S 7 we provide some limiting arguments endorsing our results from the perspective of AdS/CFT correspondence.

\bigskip \bigskip

\section{Carrollian basis for scattering amplitudes}

In his seminal work \cite{Strominger:2013jfa}, Strominger proved that the gravitational S-matrices in asymptotically flat spacetimes posses an infinite dimensional symmetry group, known as the BMS group. This group of symmetries is a diagonal subgroup of the BMS$^+ \times$ BMS$^-$, where $\pm$ refers to group of asymptotic symmetries that acts on $\mathscr{I}^\pm$ respectively. As a consequence of this infinite dimensional symmetries, the S-matrix in flat spacetime factories into a universal soft factor times a lower point amplitude, when a massless insertion approaches zero energy. This is famously known as  soft theorems in the literature. Although these theorems existed independently from the sixties \cite{PhysRev.140.B516,Low:1954kd,PhysRev.110.974}, their equivalence with the Ward identities associated to the asymptotic symmetries were uncovered only last decade (see e.g. \cite{Strominger:2017zoo}). 

\medskip

For the purpose of this paper, we shall be interested in soft graviton theorems. In this case, the S-matrix admits an universal leading and subleading soft factorisation \cite{PhysRev.140.B516,Cachazo:2014fwa}. The leading and subleading soft graviton theorems can be associated with the Ward identities of supertranslation and superrotation symmetries respectively \cite{He:2014laa,Kapec:2016jld}. There are also evidences of subsubleading soft factorisation \cite{Campiglia:2016jdj,Campiglia:2016efb}. However these are not universal and here we shall consider only upto the subleading case. The leading soft graviton theorem can be expressed as
\begin{align}
	\lim_{\omega \to 0} \omega \mathcal{S}_{n+1}(k^\mu,p^\mu_i)=\sum_{i=1}^{n}\frac{\epsilon_{\mu\nu}p_i^\mu p_i^\nu}{p_i.\hat{k}}\mathcal{S}_n(p^\mu_i) \, .
\end{align} 
Here $\mathcal{S}_n$ denotes the scattering amplitudes involving $n$ massless particles. $k^\mu$ refers to the graviton 4-momentum with energy $\omega$ and unit vector $\hat{k}^\mu$, i.e. $k^\mu=\omega \hat{k}^\mu$. $p^\mu_i$ are the momentum of other massless hard particles. $\epsilon_{\mu\nu}$ denotes the polarisation tensor. 

\medskip

\noindent Similarly the subleading piece can be extracted as 
\begin{align}
	(1+\omega\partial_{\omega})\mathcal{S}_{n+1}(k^\mu,p^\mu_i)=\sum_{i=1}^{n}\frac{\epsilon_{\mu\nu}p_i^\mu\hat{J}^{\nu\lambda}k_{\lambda}}{p_i.\hat{k}} \mathcal{S}_n(p^\mu_i),
\end{align}
where $\hat{J}^{\nu\lambda}$ is the angular momentum of the graviton.

\medskip

As addressed in the introduction, a potential candidate for a holographic dual of asymptotically flat spacetime is a codimension one Carrollian CFT, that lives on the null boundary. This proposal follows from matching of symmetries from bulk and the boundary side. Null infinity, intrinsically is a Carrollian manifold and the infinite dimensional BMS group acts as a group of local conformal symmetries on the intrinsic Carrollian structures. As the gravitational S-matrix is invariant under the BMS group, it is suggestive that the scattering amplitudes can be written as a correlation function of these Carrollian CFTs \cite{Bagchi:2022emh,Donnay:2022wvx,Bagchi:2023fbj,Salzer:2023jqv,Saha:2023hsl,Bagchi:2023cen,Mason:2023mti}. For this purpose, a new basis is required \cite{Banerjee:2018gce,Banerjee:2019prz}, which makes the Carrollian properties of S-matrix element manifest \cite{Bagchi:2022emh}. This transformation of basis is achieved by a modified Mellin transformation of the plane wave basis. The modified Mellin map \cite{Banerjee:2018gce} casts the scattering amplitudes as Carrollian correlation functions as
\begin{equation}
	\begin{split}
		\langle\Phi_{h_i,\bar{h}_1}^{\epsilon_1} (u_1,z_1,\bar{z}_1)\Phi_{h_2,\bar{h}_2}^{\epsilon_2} (u_2,z_2,\bar{z}_2)\dots &\Phi_{h_n,\bar{h}_n}^{\epsilon_n} (u_n,z_n,\bar{z}_n)\rangle \\ &= \int \prod_{i=1}^{n} d\omega_i \, \omega_i^{\Delta_i-1}e^{i\epsilon_i\omega_iu_i}\mathcal{S}_n (\omega_i,z_i,\bar{z}_i,\sigma_i) \, .
	\end{split}
\end{equation}
Here $\Phi_{h_i,\bar{h}_i}(u_i,z_i,\bar{z}_i)$ are the 3d conformal Carroll primary fields with conformal weights $h$ and $\bar{h}$ and $\mathcal{S}_n(\omega_i,z_i,\bar{z}_i,\sigma_i)$ denote n-point scattering amplitudes of massless particles with helicity $\sigma_i$ and momentum 
\begin{equation} \label{para}
	p^\mu= \omega q^\mu, \quad q^\mu=(1+z\bar{z},z+\bar{z},i(z-\bar{z}),1-z\bar{z}) \, ,
\end{equation}
where $\omega$ represents the energy of the particle, and $z$ and $\bar{z}$ represent the stereographic coordinates on the celestial sphere. On a more elementary level, it is possible to define a local Carrollian primary field on null infinity using the bulk creation/annihilation operators via \cite{Bagchi:2022emh}
\begin{equation} \label{localop}
	\Phi^{\epsilon}_{h, \bar{h}}(u,z, \bar{z}) = \int_0^\infty d\omega \,e^{i\omega\epsilon u} \omega^{\Delta - 1} a(\epsilon \omega, z, \bar{z}, \sigma).
\end{equation}
The conformal weights $h$ and $\bar{h}$ of these operators can be determined in terms of $\Delta$ and the helicity $\sigma$ as

\begin{equation}
	h + \bar{h} = \Delta \quad \text{and} \quad h - \bar{h} = \sigma.
\end{equation}
The parameter $\epsilon$ can assume $\pm 1$, depending on wheather it is a creation or annihilation operator. This integral transformation ensures that the local operator  
$\Phi^{\epsilon}_{h, \bar{h}}(u,z, \bar{z})$ transforms as a primary with respect to the conformal Carroll group. The global transformation rules can be verified by plugging in, the transformation of the $a,a^\dagger$ under the Poincare group in \eqref{localop} \cite{Banerjee:2018gce}. The reader is directed to \eqref{ward} for further details about Carroll primaries. \\

\medskip

We shall now write the soft graviton theorems in the modified Mellin or Carrollian primary basis. The leading ($S_0(z,\bar{z})$) and subleading ($S_1(z,\bar{z})$) soft modes can be extracted from the momentum space graviton operator as 
\begin{align}
	S_0^{\pm}(z,\bar{z})=\lim_{\omega \to 0} \omega a (\epsilon\omega,z,\bar{z},\sigma=\pm 2), \quad S_1^{\pm}(z,\bar{z})=(1+\omega\partial_{\omega})a (\epsilon\omega,z,\bar{z},\sigma=\pm 2) \, .
\end{align}
Here  $\pm$ refers to positive and negative helicity respectively. 
The leading soft graviton theorem is given by \cite{He:2014laa}
\begin{align}
	\lim_{\omega \to 0} \omega \mathcal{M}_{n+1}(\omega,z,\bar{z},\omega_i,z_i,\bar{z}_i)=-\Big(\sum_{i}\frac{\epsilon_i\omega_i(\bar{z}-\bar{z}_i)}{z-z_i}\Big)\mathcal{M}_n(\omega_i,z_i,\bar{z}_i,\sigma_i) \, .
\end{align}
As the scattering amplitudes can be schematically expressed as correlation functions of bulk momentum space operators, alternatively we can write as 
\begin{align} \label{leading}
	\langle S_0^-(z,\bar{z})\prod_{i=1}^{n}a_i(\epsilon_i\omega_{i},z_i,\bar{z}_i,\sigma_i) \rangle =-\Big(\sum_{i}\frac{\epsilon_i\omega_i(z-z_i)}{\bar{z}-\bar{z}_i}\Big) \langle \prod_{i=1}^{n}a_i(\epsilon_i\omega_{i},z_i,\bar{z}_i,\sigma_i) \rangle \, .
\end{align}
Furthermore, the modified Mellin transformation \eqref{softop} on the hard particles yields \cite{Banerjee:2020zlg,Banerjee:2021dlm}
\begin{align} \label{sub}
	\langle	S_0^-(z,\bar{z})\prod_{i=1}^{n}\Phi^{\epsilon_i}_{h_i,\bar{h}_i}(u_{i},z_i,\bar{z}_i,\sigma_i) \rangle =-\sum_{i} \epsilon_i \frac{z-z_i}{\bar{z}-\bar{z}_i}\partial_{u_i} \langle \prod_{i=1}^{n} \Phi^{\epsilon_i}_{h_i,\bar{h}_i}(u_{i},z_i,\bar{z}_i,\sigma_i) \rangle \, .
\end{align}
Using similar arguments, we can express the subleading soft graviton theorem in modified Mellin space as \cite{Banerjee:2020zlg} 
\begin{align}
	\langle S_1^-(z,\bar{z})\prod_{i=1}^n \Phi^{\epsilon_i}_{h_i,\bar{h}_i }(u_i,z_i,\bar{z}_i) \rangle =-\left(\sum_{i}\epsilon_i\frac{(z_i-z)^2+2(h_i+u_i\partial_{u_i})(z_i-z)\partial_{z_i}}{\bar{z}-\bar{z}_i}\right) \nonumber \\ 
	\langle \prod_{i=1}^{n} \Phi^{\epsilon_i}_{h_i,\bar{h}_i}(u_i,z_i,\bar{z}_i) \rangle \, .
\end{align}
We shall comeback to these universal soft graviton theorems in modified Mellin space, i.e.   \eqref{leading} and \eqref{sub} in section \ref{four}, where we identify these relations as the Carrollian stress tensor Ward identities.




\section{Boundary Carrollian stress tensor}
\label{sec:bdystresstensor}

In this section, we turn our attention to the intrinsic construction of boundary Carroll stress tensors. Carrollian manifolds are signature zero spacetimes \cite{Henneaux:1979vn} and naturally arise on the intrinsic geometries of null hypersurfaces.
These manifolds are endowed with a covariant degenerate 2-tensor field $h_{\gamma\beta}$ and a nowhere vanishing contravariant vector field $\tau^\alpha$ orthogonal to it.
It is possible to define the inverses via the relations 
\begin{align}
	\tau^\alpha\tau_\alpha=-1, \quad h_{\gamma\beta}h^{\gamma\alpha}=\delta^\alpha_{~\beta}, \quad \tau_\alpha h^{\alpha\gamma}=0 \, .
\end{align}
However unlike relativistic manifolds, here only $\tau^\alpha$ and $h_{\gamma\beta}$ are invariant under local Carroll transformations and defines the notion of metric in this case, whereas the inverses transform as \cite{Hartong:2015xda}
\begin{equation} \label{loca}
	\delta \tau_{\mu}=e^a_{\mu}\lambda_a, \quad \delta e^a_{~\mu}=\lambda^a_b e^b_{~\mu} \, .
\end{equation}
$\lambda_a$ and $\lambda^a_b$ parametrises local Carroll boosts and local rotation. Here we have also introduced the zwiebeins $e^a_{~\mu}$.  The tensor field $h_{\gamma\beta}$ can be expressed in terms of the these zwiebeins as $h_{\gamma\beta}=e^a_{~\gamma}e^b_{~\beta}\delta_{ab}$.

\medskip

The particular example of interest for our purpose is the null boundary of 4d asymptotically flat spacetimes. The induced metric in this case is given by the line element 
\begin{align}
	ds^2=0.du^2+l_{ij}dx^idx^i \, .
\end{align}
Here, u is the usual retarded null coordinate and $x^i$ denote the coordinates of the celestial sphere. $l_{ij}$ is the round 2-sphere metric. Carrollian structures could straightforwardly be identified from the line element as 
\begin{align}
	\tau^{\alpha}=(1,0^i), \quad h_{\gamma\beta}= \text{diag} (0,l_{ij}) \, .
\end{align}
We shall also be interested in the conformal structures on these manifold. Conformal isometry equations on the given Carrollian structures are given by \cite{Duval:2014uva}
\begin{equation} \label{Weyl}
	\mathcal{L}_{\xi}\tau^\alpha=\lambda(u,x^i)\tau^\alpha, \quad \mathcal{L}_{\xi}h_{\beta\gamma}=-2\lambda(u,x^i)h_{\beta\gamma} \, .
\end{equation}
Solution to the above equation on flat Carroll manifolds is
\begin{align} \label{kill}
	\xi=\Big(\beta(x^i)+\frac{u}{2}\partial_ig^i(x^j)\Big)\partial_u+g^i(x^j)\partial_i.
\end{align}
In the above, $\beta(x^i)$ is an arbitrary function of the spatial coordinates and parametrises supertranslations. But $g^i(x^j)$ furthermore satisfies the conformal killing equations exclusively on the spatial slice. In 3d these equations restrict $g^i(x^j)$ to be holomorphic and anti-holomorphic functions, i.e. $g^z\equiv g^z(z)$ and $g^{\bar{z}} \equiv g^{\bar{z}}(\bar{z})$. One can immediately recognise that \eqref{kill} is the projection of asymptotic killing vectors on the null boundary \cite{Strominger:2017zoo}. These generators transforms the intrinsic Carrollian structures upto a Weyl factor, thus induces a conformal structure on the null boundary.

We note here that the conformal factors of $\tau^\alpha$ and $h_{\beta\gamma}$ in general could have been related by an arbitrary integer $N$, i.e. 
\be{}
\mathcal{L}_{\xi}\tau^\alpha=\lambda(u,x^i)\tau^\alpha, \quad \mathcal{L}_{\xi}h_{\beta\gamma}=-N \lambda(u,x^i)h_{\beta\gamma} \, .
\ee
This gives rise to the $N$-conformal Carroll algebras. With the above choice of $N=2$, space and time scale in the same way in the bulk AFS and the $N=2$ conformal Carroll algebra is isomorphic to the BMS algebra (in one higher dimension) \cite{Duval:2014uva}. 

\subsection{Conformal Carroll stress tensor}
\label{ssec:carstresstensor}
We now define the stress tensor components of the boundary theory as variation of the boundary action with respect to these geometric elements \cite{Baiguera:2022lsw,Dutta:2022vkg},
\begin{align}
	\delta \mathcal{S}=\int du d^2x \, e[\tau^{\nu} \delta\tau_{\mu}+e^{~\nu}_a\delta e^a_{~\nu}]T^{\mu}_{\, \, ~ \nu} \, .
\end{align}
The intrinsic Weyl and local Carroll symmetries dictates the structures of these stress tensor components.
\medskip

We first look at the transformation of the action under local Carroll transformations \eqref{loca}. Varying we get:
\begin{align}
	\delta S= \int dud^2x \, e[(\lambda_a e^a_{~\mu}\tau^{\nu}T^{\mu}_{\, \, ~\nu})+(\lambda^a_b e^b_{~\mu}e^{~\nu}_aT^{\mu}_{\, \, ~\nu})]=0 \, .
\end{align}
On flat Carroll backgrounds this implies 
\begin{align}\label{cart}
	T^i_{~ u}=0     \quad   \text{and}    \quad   T^i_{~j}=T^j_{~i}\, .
\end{align}
Notice that in contrary to the relativistic stress tensor, its Carrollian counterpart is not symmetric. This a consequence of off diagonal nature of Carroll boost transformations. It is also useful to further decompose $T^i_{~j}$ into a trace and symmetric traceless part, i.e.
\begin{equation}\label{eq:Tij}
	T^i_{~j}=t(u,x)\delta^i_{~j}+\theta^i_{~j} (u,x^i) \, .
\end{equation}
Where $t(u,x)$ denotes the trace of the spatial part only. 
\medskip
Infinitesimal Weyl rescalings of the vielbeins are given by
\begin{align}
	\delta \tau^\beta=\lambda(u,x^i) \tau^{\beta}, \quad  \delta e^{~\beta}_a=\lambda(u,x^i) e^{~\beta}_a.
\end{align}
Under these transformations the action transforms as 
\begin{align}
	\delta S=\int dud^2x \, e\lambda(u,x^i)\left[\tau^{\nu} \tau_{\mu}+e^{~\nu}_ae^a_{~\mu}\right]T^{\mu}_{\, \, ~\nu}  = \int dud^2x \, e \lambda(u,x^i)T^{\mu}_{\, \, ~\mu} \, .
\end{align}
Thus, invariance of the action implies vanishing of trace, i.e $T^{\alpha}_{~\alpha}=0$.
Tracelessness fixes $t(u,x^i)$ to be 
\begin{equation}
	t(u,x^i)=-\frac{1}{2}T^u_{~u}.
\end{equation}

\subsection{BMS currents}
\label{ssec:bmscurrents}

We use this stress tensors to evaluate the currents associated with the BMS killing vectors in (\ref{kill}) as
\begin{equation}
	J^\mu=T^{\mu}_{\, \,~\nu}\xi^\nu
\end{equation}
Conservation of these currents follow from this specific structure of the stress tensor components and the conformal killing equations
\begin{align}
	\nabla_{\mu}J^{\mu}=&\nabla_{\mu}(T^{\mu}_{~\, \, \nu}\xi^{\nu}) = T^{\mu}_{~\, \, \nu}(\nabla_{\mu}\xi^{\nu})+\xi^\nu(\nabla_{\mu}T^{\mu}_{~\, \,\nu})=0 \, .
\end{align}
In the above expression we have used the stress tensor conservation equations and the conformal killing equations.
The corresponding charges can be obtained by integrating $J^u$ over the space slice
\begin{align} \label{blocks}
	Q_{\xi}&=\int \sqrt{q}d^2z J^u =\int \sqrt{q}d^2z \left[T^u_{~u}\xi^u+T^u_{~i}\xi^i \right]   \\ \nonumber
	&=\int \sqrt{q}d^2x^i \left[T^u_{~u} \left( \beta(x^i)+\frac{u}{2}D_ig^i(x^j) \right) +T^u_{~i}.g^{i}(x^j)\right]  \\ \nonumber
	&=\int \sqrt{q} d^2x^i \left[T^u_{~u}\beta(x^i)+\left(T^u_{~i}-\frac{u}{2}D_iT^u_{~u} \right)g^i(x^j)\right] \, .
\end{align}
To get to the last line of the above equation we have dropped a total derivative term. Now consider the stress tensor conservation equations:
\begin{align}
	\partial_uT^u_{~u}+\nabla_iT^i_{~u}=0   \implies \partial_u T^u_{~u}=0 \, . 
\end{align}
In getting to the above, we have used \refb{cart}. We also have:
\begin{align}
	\partial_u T^u_{~i}+\partial_j T^j_{~i}=0 \, , \\
	\partial_u T^u_{~i}+\partial_j \left(-\frac{1}{2}T^u_{~u}\delta^j_{~i}+\theta^j_{~i} \right) =0 \, .
\end{align}
Integration with respect to $u$ yields 
\begin{align}
	T^u_{~i}-\frac{u}{2}\partial_i T^u_{~u}= C_i(z,\bar{z})-\int du \partial_j \theta^j_{~i} (u,z,\bar{z}) \, .
\end{align}
Here $C_i(z,\bar{z})$ is the integration constant and in principle can be an arbitrary function of $z$ and $\bar{z}$. From now on, we shall use the following definitions: 
\begin{equation}\label{Tcar}
	T_u(z,\bar{z}) \equiv T^u_{~u}, \quad  T_i(z,\bar{z}) \equiv  T^u_{~i}-\frac{u}{2}\partial_iT^u_{~u}=C_i(z,\bar{z})-\int du \, \partial_j \theta^j_{~i} (u,z,\bar{z}) \, .
\end{equation} 
Using the above expressions it is possible to express the charges as 
\begin{align}\label{eq:bmscharge}
	Q_\xi=\int d^2z \left[T_u(z,\bar{z})\beta(z,\bar{z})+T_z(z,\bar{z})g^z(z)+T_{\bar{z}}(z,\bar{z})g^{\bar{z}}(\bar{z})\right] \, .
\end{align}
 To get the standard expressions, we now decompose the charges by choosing the parameters in this specific way
\bes\label{Charge standard}
\begin{align} 
	L_n=Q_{\xi}[\beta=0,g^z=z^{n+1},g^{\bar{z}}=0]&=\int d^2z \sqrt{q} \, T_z. z^{n+1} \, ,   \\ 
		\bar{L}_n=Q_{\xi}[\beta=0,g^z=0,g^{\bar{z}}=\bar{z}^{n+1}]&=\int d^2z \sqrt{q} \, T_{\bar{z}}. \bar{z}^{n+1} \, ,  \\ 
	\text{and} \quad M_{r,s}[\beta=z^r\bar{z}^s,g^z=0,g^{\bar{z}}=0]&=\int d^2z \sqrt{q} \, T_u.z^r\bar{z}^s \, .
\end{align}
\ees
Here $L_n$ and $\bar{L}_n$ are superrotation charges that generate the conformal transformations on $S^2$ and $M_{r,s}$, the supertranslation charges that generate the angle dependent translation along $u$ direction.

\subsection{Ward identities} \label{ward}

The above construction would allow us to express the standard  definition of BMS$_4$ primary fields in terms of the stress tensor OPE and subsequently the Ward identities. Let us first recall the 
transformation rules associated with these primaries \cite{Bagchi:2022emh}. A Carrollian conformal or BMS$_4$ primary field $\Phi(u,z,\bar{z})$ is labelled by the eigenvalues of $L_0$ and $\bar{L}_0$ operator, i.e,  
\begin{align}
	[L_0,\Phi(0)]=h \Phi(0) \, , \quad [\bar{L}_0,\Phi(0)]=\bar{h}\Phi(0) \, ,
\end{align}
and the primary conditions are obtained by demanding that positive modes of both supertranslation and superrotation generators annihilate the fields at origin,  
\begin{align}
	[L_n,\Phi(0)]=0, \quad [\bar{L}_n,\Phi(0)]=0 \quad \forall n> 0, \quad  [M_{r,s},\Phi(0)]=0  \quad  \forall r,s >0. 
\end{align}
 at an arbitrary spacetime point,
\begin{equation}\label{Primary}
	\begin{split}
		&[L_n,\Phi(u,z,\bar{z})]=z^{n+1}\partial_z\Phi(u,z,\bar{z})+(h+\frac{u}{2}\partial_u)\Phi(u,z,\bar{z})\partial_z(z^{n+1}) \, ,  \\  
		&[\bar{L}_n,\Phi(u,z,\bar{z})]=\bar{z}^{n+1}\partial_{\bar{z}}\Phi(u,z,\bar{z})+ (\bar{h}+\frac{u}{2}\partial_u)\Phi(u,z,\bar{z})\partial_{\bar{z}}(\bar{z}^{n+1})
		\, , \\
		&[M_{r,s},\Phi(u,z,\bar{z})]=z^r\bar{z}^s\partial_u\Phi(u,z,\bar{z}) \, .
	\end{split}
\end{equation}

 These transformation rules can be encoded in terms of the operator product expansion of the stress tensor components and the operators of concern. Using (\ref{Charge standard}) it is possible to rewrite \eqref{Primary} as
\begin{align}\label{T-Phi}
	:T_z(z,\bar{z})\Phi(u,\omega,\bar{\omega}): &\sim (h+\frac{u}{2}\partial_u)\Phi(u,\omega,\bar{\omega})\partial_{\omega}\delta^2(z-\omega)+\delta^{2}(z-\omega)\partial_{\omega}\Phi(u,\omega,\bar{\omega}) \, , \nonumber\\ 
	:T_u(z,\bar{z})\Phi(u,\omega,\bar{\omega}): &\sim \partial_u\Phi(u,\omega,\bar{\omega})\delta^2(z-\omega) \, .
\end{align}
A similar expression exist for the antiholomorphic component as well. These OPEs are generic operator statements, however when inserted inside correlation functions leads to Ward identities. The stress tensor Ward identities associated with a generic 3d Carrollian CFT are given by
\begin{align} \label{Ward id}
	&\langle T_z(z,\bar{z})\prod_{i=1}^n \Phi^{\epsilon_i}_{h_i,\bar{h}_i}(u_i,z_i,\bar{z}_i))\rangle \nonumber \\ & ~~~~=\Big(\sum_{i}(h_i+\frac{u_i}{2}\partial_{u_i})\partial_{z_i}\delta^2(z-z_i) +\delta^2(z-z_i)\partial_{z_i}\Big) \langle\prod_{i=1}^{n} \Phi^{\epsilon_i}_{h_i,\bar{h}_i}(u_i,z_i,\bar{z}_i)\rangle \, , \nonumber\\ 
	\langle
	&T_u(z,\bar{z})\prod_{i=1}^n \Phi^{\epsilon_i}_{h_i,\bar{h}_i}(u_i,z_i,\bar{z}_i)\rangle=-\Big(\sum_{i} \delta^2(z-\omega_i)\partial_{u_i}\Big)\langle \prod_{i=1}^{n} \Phi^{\epsilon_i}_{h_i,\bar{h}_i}(u_i,\omega_i,\bar{\omega}_i) \, \rangle.
\end{align}

\section{Soft theorems as stress tensor Ward identities} \label{four}

In this section, we will establish the primary result of our paper. In \S\ref{sec:bdystresstensor}, we have assembled the elements that make up an arbitrary intrinsic Carrollian stress tensor. We have identified a specific decomposition in \eqref{Tcar} whose insertions in correlation functions result in Ward identities of the form \eqref{Ward id}. We will now show that these Ward identities of the boundary Carrollian stress tensor are manifestations of the bulk soft graviton theorems. We shall use the standard parameterisation of massless particles \eqref{para}. Using these parameters the leading soft graviton theorem with negative helicity \eqref{gravisoft} takes the following form
\begin{align}
	\lim_{\omega \to 0} \omega \mathcal{M}_{n+1}(\omega,z,\bar{z},\omega_i,z_i,\bar{z}_i)=-\Big(\sum_{i}\frac{\epsilon_i\omega_i(z-z_i)}{\bar{z}-\bar{z}_i}\Big)\mathcal{M}_n(\omega_i,z_i,\bar{z}_i,\sigma_i) \, .
\end{align}
After performing modified Mellin transformation on the hard particles, this relation becomes \cite{Banerjee:2020zlg}
\begin{align}
	\langle S_0^-(z,\bar{z})\prod_{i=1}^{n} \Phi^{\epsilon_i}_{ h_i,\bar{h}_i}(u_i,z_i,\bar{z}_i)\rangle =-\Big(\sum_{i}\frac{\epsilon_i(z-z_i)}{\bar{z}-\bar{z}_i}\partial_{u_i}\Big)\langle \prod_{i=1}^{n} \Phi^{\epsilon_i}_{h_i,\bar{h}_i}(u_i,z_i,\bar{z}_i)\rangle \, .
\end{align}
Acting twice with $\partial_{z}$ on either side of the above end up with 
\begin{align}
	\langle \partial^2_{z}S_0^-(z,\bar{z})\prod_{i=1}^{n} \Phi^{\epsilon_i}_{h_i,\bar{h}_i}(u_i,z_i,\bar{z}_i)\rangle=-\sum_{i} \delta^2(z-z_i) \epsilon_i\partial_{u_i}\langle\prod_{i=1}^{n} \Phi^{\epsilon_i}_{h_i,\bar{h}_i}(u_i,z_i,\bar{z}_i)\rangle \, .
\end{align}
The above relation can be recognised as Ward identity of $T_u(z,\bar{z})$ if we identify,
\begin{align}
	T_u=\partial^2_{z}S_0^-(z,\bar{z}) \, .
\end{align}
Similarly, for the positive helicity soft graviton, we have
\begin{align}
	\lim_{\omega \to 0} \omega \mathcal{M}_{n+1}(\omega,z,\bar{z},\omega_i,z_i,\bar{z}_i)=-\Big(\sum_{i}\frac{\epsilon_i\omega_i(\bar{z}-\bar{z}_i)}{z-z_i}\Big)\mathcal{M}_n(\omega_i,z_i,\bar{z}_i,\sigma_i) \, .
\end{align}
This would lead to the following identification:
\begin{equation}
	T_u=\partial^2_{\bar{z}}S_0^+(z,\bar{z}).
\end{equation}
So, in conclusion, we have a relation between the positive/negative helicity soft graviton and the $(uu)$ component of the 3d Carroll stress tensor:
\begin{align}
	\boxed{T_u(z, \z)=\partial^2_{z}S_0^-(z,\bar{z})=\partial^2_{\bar{z}}S_0^+(z,\bar{z}).}
\end{align}

\medskip

Analogously, the subleading soft graviton theorem in modified Mellin space can be identified with the Ward identities of $T_z$. The subleading soft graviton theorem in modified Mellin space is given by \cite{Banerjee:2020zlg}: 
\begin{align}
	\langle S_1^{-}(z,\bar{z})\prod_{i=1}^n \Phi^{\epsilon_i}_{h_i,\bar{h}_i }(u_i,z_i,\bar{z}_i) \rangle=-\Big(\sum_{i}\epsilon_i\frac{(z_i-z)^2+2(h_i+u_i\partial_{u_i})(z_i-z)\partial_{z_i}}{\bar{z}-\bar{z}_i}\Big) \\ \nonumber
	\langle \prod_{i=1}^{n} \Phi^{\epsilon_i}_{h_i,\bar{h}_i}(u_i,z_i,\bar{z}_i)\rangle \, .
\end{align}
Here, we denote a negative helicity subleading soft graviton by $S^-_1(z,\bar{z})$. This time we need to act with $\partial^3_z$ on both sides of the equation. This would result to 
\begin{align}
	\langle \partial^3_zS_1^{-}(z,\bar{z})\prod_{i=1}^n \Phi^{\epsilon_i}_{h_i,\bar{h}_i }(u_i,z_i,\bar{z}_i) \rangle=-\Big(\sum_{i} \epsilon_i \left(h_i \right. & \left.+\frac{u_i}{2}\partial_{u_i}\right)\partial_{z_i}\delta^2(z-z_i)  \\\nonumber
	& \quad+\delta^2(z-z_i) \epsilon_i\partial_{z_i}\Big)\langle \prod_{i=1}^{n} \Phi^{\epsilon_i}_{h_i,\bar{h}_i}(u_i,z_i,\bar{z}_i)\rangle .
\end{align}
This Ward identity is identical with \eqref{Ward id} if we have
\begin{align}
	\boxed{T_z(z, \z)=\partial^3_zS_1^{-}(z,\bar{z}).}
\end{align}
Similarly using the positive helicity subleading soft graviton theorem, one can verify that 
\begin{align}
	\boxed{T_{\bar{z}}(z, \z)=\partial^3_{\bar{z}}S_1^{+}(z,\bar{z}).}
\end{align}
So, in conclusion, we have seen that the objects $\{T_u, T_z, T_\z\}$ built out of components of the boundary 3d Carroll stress tensor are connected to the leading and subleading soft graviton operators in the 4d asymptotically flat bulk spacetime and the Carroll stress tensor Ward identities can be recast as soft theorems. 

\section{Celestial and Carrollian soft operators}\label{soft}

The purpose of this section is to elaborate on the relation between soft sectors in Celestial and Carrollian holography. We shall first briefly discuss elementary aspects of celestial holography emphasising on the soft operators, before moving on to Carrollian holography and then establish the relation between the 3d Carroll stress tensor and its 2d Celestial counterparts. 

\subsection{Celestial soft operators}
As elaborated before, Celestial holography proposes a 2d CFT dual to gravity in 4d asymptotically flat spacetimes. The basic proposal of the celestial holography emerges from the observation that Lorentz group acts as a group of global conformal symmetries on the celestial sphere of null infinity.

The existence of the subleading soft theorem demands that the global conformal group of the Celestial CFT should be extended to include all the local conformal transformations as well, thus realising the full infinite dimensional Virasoro algebra. Guided by this observation it was proposed by Strominger that gravitational scattering in asymptotically flat spacetimes is dual to a 2d CFT that lives on the celestial sphere of $\mathscr{I}^\pm$.

\medskip

In order to motivate this correspondence a new basis is introduced where the boost generators are diagonalized instead of the translations \cite{Pasterski:2016qvg,Pasterski:2017kqt}. This  basis, known as the conformal primary basis is related to the usual plane wave basis by a Mellin transformation with respect to the energy variable. After this Mellin transformation, the S-matrix elements in flat space take the form of conformal primary correlation function of a 2d CFT, i.e.
\begin{align} \label{scattering}
	\langle\mathcal{O}_{1}(z_1,\bar{z}_1)\mathcal{O}_2(z_2,\bar{z}_2)...\mathcal{O}_n(z_n,\bar{z}_n)\rangle\sim \int_{0}^{\infty}\prod_{i=1}^n \omega_{i}^{\Delta_i-1}\mathcal{S}_n(\omega_i,z_i,\bar{z}_i,\sigma_i).
\end{align}
This is the principal holographic map in celestial approach. Here $\mathcal{S}_n(\omega_i,z_i,\bar{z}_i,\sigma_i)$ denotes the $n$-point scattering amplitudes of massless particles with momentum $p_i^\mu$ and helicities $\sigma_i$ and $\mathcal{O}(z,\bar{z})$  denotes primary operators on celestial sphere\footnote{A slight modification of this framework can incorporate the massive scattering amplitudes as well. For our purpose, i.e addressing soft theorems, \eqref{scattering} suffices. We refer to \cite{Pasterski:2016qvg,Iacobacci:2022yjo} for the massive case.}. 
In the above expression the null momenta are expressed in the standard parametrisation, given by \eqref{para}. The conformal weights $(h,\bar{h})$ of these primary operators $\mathcal{O}(z,\bar{z})$ can be expressed in terms of the boost weights and helicities of the bulk particles as 
\begin{align}
	h=\frac{\Delta+\sigma}{2} \quad \text{and} \quad  \bar{h}=\frac{\Delta-\sigma}{2}.
\end{align}

\medskip

Let us now try to understand how the soft limits are realised in the Mellin space. In Mellin space, energy is traded with the conformal weights $\Delta$ of the operators. It turns out that via the celestial map \eqref{scattering}, the poles in the soft expansion, are mapped to the poles in the conformal dimension $\Delta$ of the associated primary operators. For example, one can extract the leading and subleading soft contribution $G_0(z,\bar{z})$ and $G_1(z,\bar{z})$  from a  graviton conformal primary operator $G_{\Delta}(z,\bar{z})$ as  \cite{Donnay:2018neh}
\begin{equation}
	S^+_0(z,\bar{z})=\text{Res}_{\Delta=1}G_{\Delta,\sigma=2}(z,\bar{z}), \quad S^+_1(z,\bar{z})=\text{Res}_{\Delta=0}G_{\Delta,\sigma=2}(z,\bar{z}).
\end{equation}
Using these leading and subleading graviton conformal primaries, one can define the following operators of the dual CFT \cite{Strominger:2013jfa,Barnich:2013axa,He:2014laa,Strominger:2014pwa,Donnay:2018neh}
\bes\label{softop}
\begin{align} 
	& P_{(2d)}(z,\bar{z})=\partial_{\bar{z}}S^+_0(z,\bar{z}), \\
	& \bar{T}_{(2d)}(\bar{z})=\int d^2w \frac{1}{(\bar{z}-\bar{w})}\partial_{\bar{w}}^3S^+_1(w,\bar{w}). 
\end{align}
\ees
The integral transformation in the later expression is called the Shadow transformation \cite{Donnay:2018neh}. One peculiarity of this transformation is that it maps a primary operator of holomorphic weight $(h,\bar{h})$ to another one of $(1-h,1-\bar{h})$. Notice that the subleading soft graviton $S^+_1(z,\bar{z})$ has weights $(1,-1)$. Thus after the shadow transformation it gives rise to another primary $\bar{T}_{(2d)}(\bar{z})$ of weight $(0,2)$. The weights of the operator $P_{(2d)}(z,\bar{z})$ can similarly counted to be $(\frac{3}{2},\frac{1}{2})$.

\medskip

Now the leading and subleading soft factorisation would lead to the Ward identities of these 2d CFT operators we defined in \eqref{softop}. It is possible to show by plugging in the leading soft graviton theorem \eqref{leading} in the holographic map  \eqref{scattering}, the insertion of $P(z,\bar{z})$ in the correlation functions leads to the following Ward identity \cite{He:2014laa}
\begin{equation} \label{celestialkac}
	\langle P_{(2d)}(z,\bar{z})\prod_{i=1}^{n} \Phi_{h_i,\bar{h}_i }(w_i,\bar{w}_i) \rangle =\sum_{i=1}^{n}\frac{1}{z-w_i}	\langle \prod_{i=1}^{n} \Phi_{h_i+\frac{1}{2},\bar{h}_i +\frac{1}{2}}(w_i,\bar{w}_i) \rangle.
\end{equation}
Subsequently, from the subleading soft theorem,  we shall have 
\begin{equation} \label{Shadow}
	\langle \bar{T}_{(2d)}(\bar{z})\prod_{i=1}^{n} \Phi_{h_i,\bar{h}_i }(w_i,\bar{w}_i) \rangle= \sum_{i=1}^{n}\Big(\frac{\bar{h}_i}{(\bar{z}-\bar{w}_i)^2}+\frac{1}{\bar{z}-\bar{w}_i}\partial_{\bar{w}_i}\Big)\langle \prod_{i=1}^{n} \Phi_{h_i,\bar{h}_i }(w_i,\bar{w}_i) \rangle.
\end{equation}
Thus the subleading soft theorem implies the existence of a Virasoro stress tensor in Celestial CFT \cite{Kapec:2016jld}. {\footnote{A generalisation for higher dimensions was constructed in \cite{Kapec:2017gsg}.} This stress tensor defined in \eqref{softop} is known as the Shadow stress tensor. Similarly the leading soft graviton theorem demands that the dual Celestial CFT must contain a current of weight $(\frac{3}{2},\frac{1}{2})$ \cite{Donnay:2018neh}. 

\subsection{Carrollian stress tensors}
\label{ssec:cartensor}

Here we return to our construction of Carrollian stress tensor and establish its relation to the celestial soft operators. We remind the reader that there are three stress tensor components $T_u(z,\bar{z}),T_z(z,\bar{z}),T_{\bar{z}}(z,\bar{z})$, that contributes to the BMS$_4$ charges in \eqref{eq:bmscharge}. We shall use the algebra of these charges to figure out the transformation rules of these components. We finally use the results of this section to relate to the carroll stress tensor to the celestial operators defined in \eqref{softop}. 

\medskip

The standard form of the non-central BMS$_4$ charge algebra is given by \eqref{bms4}. Using the expressions of charges in \eqref{Charge standard}, it is possible to show that the stress tensor components $T_z(z,\bar{z}),T_{\bar{z}}(z,\bar{z}), T_u(z,\bar{z})$ transforms as primaries with respect to the Virasoro subgroup of BMS$_4$ \cite{Dutta:2022vkg}. Below we determine the holomorphic weights of these primary operators. For example if we consider the commutation relation
\begin{align}
	[L_n,L_m]=(n-m)L_{n+m},
\end{align}
this immediately fixes the transformation rules of $T_z$ with respect to the left moving Virasoro generators:
\begin{align} 
	[L_n,T_z(z,\bar{z})]=z^{n+1}\partial_z T_z(z,\bar{z})+(n+1)2T(z,\bar{z})z^n. 
\end{align}
Similarly from
\begin{align}
	[\bar{L}_m,L_n]=0
\end{align}
we have 
\begin{align}
	[\bar{L}_n,T_z(z,\bar{z})]=\bar{z}^{n+1}\partial_{\bar{z}}T_z(z,\bar{z})+(n+1)T_z(z,\bar{z})\bar{z}^n.
\end{align}
Using similar methods it is straightforward to show that $T_u$ and $T_{\bar{z}}$ also transforms as primary fields. The conformal weights of these operators are listed below in Table~\ref{t1}.
\begin{table}[h!]
	\centering
	\begin{tabular}{||c | c c c||} 
		\hline
		weights &  $T_u$ & $T_z$ & $T_{\bar{z}}$ \\ [0.75ex] 
		\hline\hline
		$\Delta$ & $3$ & 3 & 3 \\
		\hline 
		$h$ &  ${3}/{2}$ & 2 & 1 \\ 
		\hline
		$\bar{h}$ & ${3}/{2}$ & 1 & 2 \\
		\hline

	\end{tabular}
	\caption{Weights of Carroll stress tensor components}
	\label{t1}
\end{table}

Using the commutator between $L_n$ and $M_{rs}$, it is also possible to fix transformation property of these objects with respect to supertranslations. These are
\bes
\begin{align}
	\delta_{\beta(z,\bar{z})}T_z=& \, \frac{1}{2}\beta(z,\bar{z}) \partial_z T_u(z,\bar{z})+\frac{3}{2}T_u(z,\bar{z})\partial_z\beta(z,\bar{z}) \, , \\ 
	\delta_{\beta(z,\bar{z})}T_{\bar{z}}=& \, \frac{1}{2}\beta(z,\bar{z}) \partial_{\bar{z}} T_u(z,\bar{z})+\frac{3}{2}T_u(z,\bar{z})\partial_{\bar{z}}\beta(z,\bar{z}) \, , \\
	\delta_{\beta(z,\bar{z})}T_u=& \,0 \, .
\end{align}
\ees
We note here that while 3d Carroll stress tensor $T^a_{\,\,b}(u, z, \z)$ is a primary with respect to the 3d Conformal Carroll algebra, the components $\{T_z(z,\bar{z}),T_{\bar{z}}(z,\bar{z}) \}$ transform as primaries with respect to only the Virasoro sub-algebra of the Conformal Carroll algebra. {\footnote{$T_u(z, \z)$ is a primary of the full Conformal Carroll algebra, and hence is of course a primary wrt the Virasoro subalgebras.}}

\subsection{Relation between the soft sectors}
\label{ssec:rel}

A series of seminal papers \cite{Strominger:2013jfa,He:2014laa,Strominger:2014pwa,Cachazo:2014fwa,Kapec:2016jld}, it was shown that the soft theorems are Ward identities of the asymptotic symmetries in flat spacetimes. This correspondence readily suggests that soft modes of the massless bulk fields can be expressed in terms of the charges corresponding to the asymptotic symmetry group \cite{Donnay:2018neh}. The two different proposals of putative holographic dual, i.e celestial and Carrollian, emerges depending on how this symmetry group is packaged into a field theory as a group of global symmetries. Equating the charges on both side of the duality leads us directly to a relation between the soft modes and the dual field theory operators. 

\medskip

In Celestial framework, this relation is given by \eqref{softop}. The stress tensor in a celestial CFT is defined as a shadow transformation of the subleading soft graviton operator. As a consequence of this integral transformation the stress tensor is a non-local operator in the dual field theory \cite{Banerjee:2022wht}. In addition to Virasoro stress tensor, a $U(1)$ Kac-Moody current is also required. This Kac-Moody current is defined using the leading soft graviton and generates the supetranslation in dual field theory. The conformal weights $(h,\bar{h})$ of $P_{(2d)}(z,\bar{z})$ and $\bar{T}_{(2d)}(\bar{z})$ are $(\frac{3}{2},\frac{1}{2})$ and $(0,2)$ respectively.

\medskip

On the other hand, in Carrollian framework the soft sector is entirely contained in the stress tensors. In \eqref{Tcar}, we have shown that it is possible to define the primary operators $T_u(z,\bar{z}), T_z(z,\bar{z})$ and $T_{\bar{z}}(z,\bar{z})$ as a combination of the Carrollian stress tensor components. These objects have the conformal weights $(\frac{3}{2},\frac{3}{2}),(2,1)$ and $(1,2)$ respectively as per Table \ref{t1}. We established the relation between these components and the soft gravitons in \S\ref{four} by comparing their the Ward identities and the soft theorems in modified Mellin space. From all these considerations it is immediately suggestive that celestial and Carrollian soft operators for the positive helicity graviton are related via the following relations
\begin{align}
	T_u(z,\bar{z})=\partial_{\bar{z}}P_{(2d)}(z,\bar{z}), \quad T_z(z,\bar{z})=\partial_{\bar{z}}T_{(2d)}(z), \quad T_{\bar{z}}(z,\bar{z})=\partial_z\bar{T}_{(2d)}(\bar{z}).
\end{align}
The celestial stress tensor $\bar{T}_{(2d)}(\bar{z})$ is a non-local object. Hence the action of $\partial_{z}$ on $\bar{T}_{(2d)}(\bar{z})$ does not vanish. On the contrary, it localizes the shadow integral giving rise to another local primary of weight $(1,2)$, i.e.
\begin{align}
	T_{\bar{z}}(z,\bar{z})&=\partial_z\bar{T}_{(2d)}(\bar{z})=\partial_{z}\int d^2w \, \frac{1}{\bar{z}-\bar{w}}\partial_{\bar{w}}^3S^+_1(w,\bar{w}) \\ \nonumber
	&=\int d^2w \, \, \delta^{2}(z-w)\partial^3_{\bar{w}}S_1(w,\bar{w}) =\partial^3_{\bar{z}} S^+_1(z,\bar{z}) \, .
\end{align}
This primary operator naturally arises in the stress tensor sector of a generic 3d Carroll CFT. A similar result holds for the holomorphic $T_{z}(z,\bar{z})$ which also localises the non-local 2d holomorphic stress tensor. 

\section{Stress tensor from a limit and connections to AdS/CFT}\label{limit}
We have seen above that in contrast to the two dimensional proposed Celestial dual theory where one needs the Celestial stress tensor and an additional current to connect to the leading and subleading soft graviton theorems, the 3d Carrollian theory packages both soft theorems into a single stress tensor with the $(uu)$ component connecting to the leading and a combination of the $(ui)$ and $(uu)$ components identified to the sub-leading soft theorems. In this section, we offer some explanations as to why one should expect this to happen and outline a process of connecting this to usual holographic ideas from AdS/CFT. 

\subsection{A simple example: scalar field}
We will begin with the simple example of a relativistic scalar in $d=3$ and its relation with the 3d Carroll scalar. The Noetherian stress tensor for a massless relativistic scalar $\phi$ in Minkowski $d=3$ spacetime (with signature $(-++)$ ) is given by
\begin{equation}
	T^{\mu}_{~~\nu \, \text{(old)}} = \partial^{\mu} \phi \partial_{\nu} \phi - \dfrac{1}{2}\delta^{\mu}_{~\nu}(\partial_{\rho}\phi \partial^{\rho} \phi) \, ,
\end{equation}
where the Greek indices $\mu,\nu,\rho$ run over the spacetime coordinates $(t,x^i)$. The above stress tensor is not traceless, thus, we do the standard Belinfante improvement to obtain 
\begin{equation}
	T^{\mu}_{~~\nu} = T^{\mu}_{~\nu \, \text{(old)}} - \dfrac{1}{8}\left(- \delta^{\mu}_{~\nu} \Box + \partial^{\mu} \partial_{\nu} \right) \phi^2 \, .
\end{equation}
We have $T^{\mu}_{~~\mu} = 0$ on-shell. The various components are given by
\bes
\begin{align}
	T^t_{~t} &= -\dfrac{1}{2}(\partial_t \phi)^2 - \dfrac{3}{4}(\partial_i \phi)^2 - \dfrac{1}{4} \phi \partial^2_i \phi \, , \\
	T^t_{~i} &= -\dfrac{3}{4} \partial_t \phi \partial_i \phi + \dfrac{1}{4} \phi \partial_t \partial_i \phi \, , \\
	T^i_{~j} &= \dfrac{1}{4} \delta^i_{~j} (\partial_t \phi)^2 + \dfrac{3}{4} \partial_i \phi \partial_j \phi - \dfrac{1}{4} \phi \partial_i \partial_j \phi - \dfrac{1}{4} \delta^i_{~j} \left(  \partial_k \phi \right)^2 \, .
	\end{align}
	\ees
We now implement the standard Carrollian limit $t \to \epsilon \, t$, $x^i \to x^i$ with $\epsilon \to 0$. We end up with the following components:
\bes	\begin{align}
		T^t_{~t} &= -\dfrac{1}{\epsilon^2} \dfrac{1}{2} (\partial_t \phi)^2 \, \\
		T^t_{~i} &= -\dfrac{1}{\epsilon} \left(\dfrac{3}{4} \partial_t \phi \partial_i \phi - \dfrac{1}{4} \phi \partial_t \partial_i \phi \right) \, , \\
		T^i_{~j} &= - \dfrac{1}{2} T^t_{~t} \delta^i_{~j} + \mathcal{O}(\epsilon^0) \, .
	\end{align}
\ees
At the moment, we focus on $T^t_{~t}$ and $T^t_{~i}$. We will return to $T^i_{~j}$ near the end of this section. 
We redefine:
\be{}
T^u_{~u} = \e^2 \, T^t_{~t} = - \dfrac{1}{2} (\partial_u \phi)^2, \quad T^u_{~i} =\e  \, T^t_{~i} = -\dfrac{3}{4} \partial_u \phi \partial_i \phi +\dfrac{1}{4} \phi \partial_u \partial_i \phi.
\ee
The above are identical to the ones obtained by an intrinsic Carrollian analysis starting out with the electric Carroll scalar action and implementing the Noether procedure \cite{Baiguera:2022lsw,Dutta:2022vkg}. From the above, we see that $T^u_{~u}$ is the leading piece and $T^u_{~i}$ is next-to-leading piece in terms of $\e$ expansion which is the same as a $c$ expansion in this case.  The above scalings of the different components of the stress tensor with $c$ in the Carroll limit is a generic feature of Electric Carrollian field theories, which emerge as the leading piece in the $c\to0$ expansion. 

\subsection{Comments on limit from AdS/CFT}
Let us now return to holography. In standard AdS/CFT, it is of course well known that the bulk metric is dual to the stress tensor on the boundary. So the fluctuations of the bulk metric or gravitons are captured by the correlations of the boundary stress tensor. The flat limit of AdS/CFT can be viewed as a very high energy limit where one focuses on a tiny Minkowski diamond at the centre of AdS. This very small region is unable to perceive the curvature of AdS. This is in keeping with the idea that a Carrollian CFT can be thought of as a very high energy limit of a usual relativistic CFT. This e.g. has been explicitly demonstrated by an infinite boost of a relativistic CFT to a Carroll CFT in $d=2$ \cite{Bagchi:2022nvj}. So a flat space limit in the bulk which is Carroll limit on the boundary focuses only on the very high energy modes of the original theory. Any finite energy solution in this limit would thus naturally scale to zero energy and become soft. 

We can understand this from a different perspective as well. The typical large AdS radius limit of Penedones' formula \cite{Penedones:2010ue}, the flat space scattering amplitude is encoded in the limit when the Mandelstam variables become large. Thus, if we do not scale the energy of the scattering process in the bulk of AdS with the AdS radius, we end up with particles going soft in the flat limit. This is roughly due to the wavelength being stretched by an order of the AdS radius as depicted in figure \ref{fig:scattering}. 

\begin{figure}[h]
	\centering
	\includegraphics[width=11cm, height=8cm]{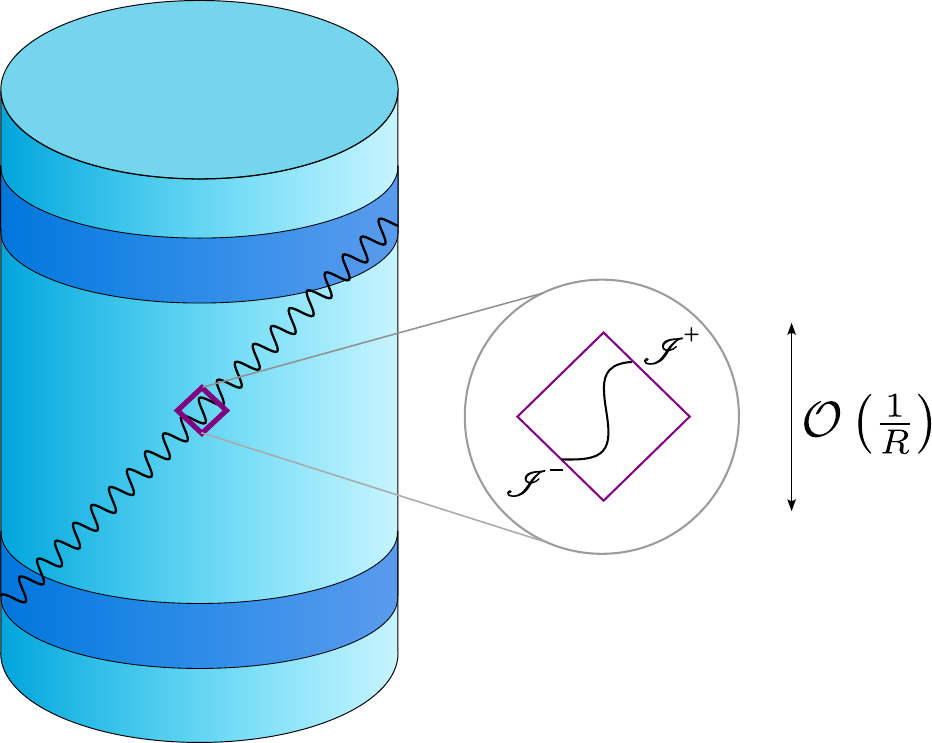}
	\caption{Wavelengths of bulk scattering processes in AdS get stretched out as we zoom into the Minkowski diamond at the centre of AdS.}
	\label{fig:scattering}
\end{figure}

The mode of the graviton corresponding to the stress tensor component $T^u_{\, u}$ goes soft fastest because this is the leading term in the $c\to0$ limit. This is what accounts for the leading Weinberg soft theorem in 4d AFS. $T^u_{\, u}$ also generate supertranslations (as per \refb{Charge standard}). We thus have an understanding of why supertranslation Ward identities lead to the leading soft theorem purely from the point of view of the limit.  

$T^u_{\, i}$ is subleading in powers of $c$ as compared to $T^u_{\, u}$. The combination of these components that we have been interested in is 
\be{}
T_z = T^u_{\, z} - \frac{1}{2} u \, \p_i T^u_{\, u}.
\ee
The two terms in $T_z$ scale the same way with $c$, since $u$ scales as $c$ and $x_i$ does not scale. And this is subleading compared to $T^u_{\, u}$, as can be seen from the explicit scalar example as well as from the fact that the combination above has a $u$ multiplying the $T^u_{\, u}$ term. The modes of the graviton corresponding to this combination of stress tensor components thus goes soft slower than the modes of the graviton corresponding to $T^u_{\, u}$. This combination also generates superrotations and corresponds to the sub-leading soft graviton theorem.  

The components of the Carroll stress tensor thus naturally capture the soft sector of gravitational radiation in the bulk 4d asymptotically flat spacetime.  From the point of view of a limit of AdS/CFT, this is because the finite energy gravitons are stretched out in the infinite radius limit which corresponds to the zero lightspeed limit in the dual CFT. 

\subsection{Hard scattering?}
An immediate open direction would be to understand the precise relation between the scattering of hard gravitons and our stress tensor. Our analysis has so far been symmetry based and hence it is interesting to see how scattering fits in. Keeping this in mind let us comeback to \eqref{eq:Tij} and have a closer look at the $\theta^i_{~j}$ term. As introduced in section \ref{sec:bdystresstensor}, $\theta^i_{~j}$ is the symmetric traceless part of spatial stress tensor components: 
\begin{equation}
	\theta^i_i=0, \quad \theta^i_j=\theta^j_i.
\end{equation}
These two condition leaves us with two independent components, which in $z,\bar{z}$ coordinates can be expressed as $\theta_{zz}$ and $\theta_{\bar{z}\bar{z}}$. Now it is straightforward to read off how these components scale (i.e. their weights under the action of $L_0$ and $\bar{L}_0$) directly from the stress tensor conservation equations. For example the $z$-component of \eqref{Tcar} reads 
\begin{align}
	T_z(z,\bar{z})=C_z(z,\bar{z})-\int du d^2z \partial_{z}\theta_{zz} (u,z,\bar{z}).
\end{align}
 $T_z(z,\bar{z})$ transform as a Virasoro primary of weight (2,1). Also the coordinates ($u,z,\bar{z}$) scale with weights ($-\frac{1}{2},-\frac{1}{2}$),$(-1,0)$ and $(0,-1)$ respectively. Thus $\theta_{zz}(u,z,\bar{z})$ must have scaling dimensions ($\frac{5}{2},\frac{1}{2}$). 
 
 Similarly from the $\bar{z}$ component of \eqref{Tcar}, one can verify that $\theta_{\bar{z}\bar{z}}(u,z,\bar{z})$ transforms with weights ($\frac{1}{2},\frac{5}{2}$). It is evident that $\theta^i_j(u,z,\bar{z})$ terms comprise the pure spin 2 sector of the Carrollian stress tensor. Now as the modified Mellin map relates spin of the boundary operators with the helicty of the massless particles involved in the bulk scattering process, it is tempting to consider that the correlation functions of these $\theta^i_j(u,z,\bar{z})$s would capture the hard scattering amplitudes from boundary perspective. 
 
However, the precise relation to the spin 2 primary fields we constructed from the flat space limit in \cite{Bagchi:2023cen} remains to be worked out. It is instructive to understand the set of boundary conditions in AdS \cite{Barnich:2012aw,Compere:2019bua,Compere:2020lrt} and a suitable variational principle that reduces to our stress tensor in the flat space limit. This would firmly establish an important result in the holographic dictionary: the relation between the bulk metric degrees of freedom and the boundary stress tensor similar to AdS holography {\footnote{The recent work \cite{Ciambelli:2024kre} has some interesting comments in this direction.}}.

\section{Conclusions}

We conclude this paper with a summary of our results. We have shown that the universal leading and sub-leading soft graviton theorems in asymptotically flat spacetimes can be recast as Ward identities of the intrinsic Carrollian stress tensor defined at null infinity. Any Carrollian CFT inherits a stress tensor which can be obtained from the variation of geometric elements as in \S\ref{ssec:carstresstensor}. Then in \S\ref{ssec:bmscurrents}, we shown that we can use the stress tensor to construct conserved BMS charges involving specific decompositions of the stress tensor components. Further, these decompositions lead to Ward identities in \S\ref{ward}. Finally, in \S\ref{four} we show that the modified Mellin transformation of the leading and sub-leading soft graviton theorems can be recast as Carrollian stress tensor Ward identities provided we can correctly identify the soft insertions with appropriate stress tensor components. The matching between the soft operators in the bulk and the boundary stress tensor constitutes an important step in establishing a co-dimension one holographic correspondence. By working with the intrinsic geometry of the boundary theory, we have shown that the Ward identities involving the boundary stress tensor naturally encode bulk soft theorems. Our procedure does not necessitate the inclusion of arbitrary sources to the boundary Carrollian field theory (for example \cite{Donnay:2022aba,Donnay:2022wvx}) since soft theorems are purely a consequence of symmetries. In \S\ref{soft}, we established a relation between the 3d Carroll stress tensor and the operators that determine the soft section in 2d Celestial CFT and shown in  \S\ref{ssec:rel} that the local 3d stress tensor we derived eliminates the need to introduce non-local 2d objects in the Celestial approach. In \S\ref{limit}, we indicated how the Carroll stress tensor could be obtained through the $c\to0$ limit of a 3d relativistic CFT, by considering the simple example of a massless scalar and made some comments on the connections to AdS/CFT. 

\medskip

Finally, we mention a couple of pressing directions for future work, apart from the connections to the flat limit of AdS/CFT that we have discussed in the previous section and which is perhaps one of the most important problems. 

\noindent The case of 3d AFS and 2d Carroll CFTs has been more studied in previous literature since the lower dimension makes calculations more tractable and there is the absence of gravitational radiation (in Einstein theory). The matching of stress tensor correlation functions between boundary and variations of the gravitational action was shown to arbitrary point functions in \cite{Bagchi:2015wna}. It would be of interest to revisit this analysis, since the form of OPEs used there look more like ones from magnetic Carroll theories instead of Electric ones we have used in this paper. The new work on light transformations relating these two branches \cite{Banerjee:2024hvb} might help in this regard. 

\noindent One of the outstanding questions in the fields is figuring out how double soft limits work. For this one needs to understand the stress tensor OPEs in the dual theory. Some preliminary work has been done in \cite{Dutta:2022vkg}. This is something we will return to in the near future.

\bigskip

\subsection*{Acknowledgements} 
We wish to thank Rudranil Basu, Jelle Hartong, Alok Laddha, Arthur Lipstein and Prahar Mitra for useful discussions. 

\medskip

\noindent AB is partially supported by a Swarnajayanti Fellowship from the Science and Engineering Research Board
(SERB) under grant SB/SJF/2019-20/08 and also by SERB grant CRG/2022/006165. AB and SD thanks the participants and organisers of the workshop “Carrollian Physics and Holography” organised at the Erwin Schr\"{o}dinger Institute (ESI), University of Vienna, for interesting discussions, and the ESI for hospitality during the visit. AB also thanks ULB Brussels and TU Wien for hospitality during the period of this work. PD would like to duly acknowledge the Council of Scientific and Industrial Research (CSIR), New Delhi for financial assistance through the Senior Research Fellowship (SRF) scheme. SD acknowledges support from Fonds de la Recherche Scientifique FRS-FNRS (Belgium) through the PDR/OL
C62/5 project “Black hole horizons: away from conformality” (2022-2025) and IISN – Belgium (convention 4.4503.15). SD also thanks hospitality of IIT Kanpur at early stages of work.


\medskip


\bibliographystyle{JHEP}
\bibliography{flat}

\end{document}